\documentclass[a4paper,11pt]{article}
\pdfoutput=1 

\usepackage{jheppub} 

\usepackage[utf8]{inputenc}
\usepackage{graphicx}
\usepackage{dcolumn}
\usepackage{bm}

\renewcommand{\figurename}{{\bf Figure}}
\renewcommand{\thefigure}{{\bf\arabic{figure}}}
\renewcommand{\tablename}{{\bf Table}}
\renewcommand{\thetable}{{\bf\arabic{table}}}

\usepackage{natbib}
\usepackage{slashed}
\newcolumntype{x}[1]{>{\centering\arraybackslash\hspace{0pt}}p{#1}}
\usepackage{makecell}
\def\MeV{\,\text{MeV}}
\def\GeV{\,\text{GeV}}
\def\TeV{\,\text{TeV}}

\def\mO{\mathcal{O}}
\def\mN{\mathcal{N}}
\def\MPl{M_{\text{Pl}}}
\def\ann{\text{ann}}
\def\bh{\text{bh}}
\def\gc{\text{gc}}

\def\ev{\text{ev}}

\def\cl{\text{cl}}
\def\gND{{g\text{ND}}}
\def\ND{\text{ND}}

\title{ \LARGE Magnetic monopole meets primordial black hole: an extended analysis}

\author[a]{Chen Zhang}\note{Corresponding author.}
\author[a,b,c]{and Xin Zhang}\note{Corresponding author.}

\affiliation[a]{Key Laboratory of Cosmology and Astrophysics (Liaoning) \& College of Sciences, Northeastern University, Shenyang 110819, China}
\affiliation[b]{Key Laboratory of Data Analytics and Optimization for Smart Industry (Ministry of Education), Northeastern University, Shenyang 110819, China}
\affiliation[c]{National Frontiers Science Center for Industrial Intelligence and Systems Optimization, Northeastern University, Shenyang 110819, China}

\emailAdd{zhangchen2@mail.neu.edu.cn}
\emailAdd{zhangxin@mail.neu.edu.cn}

\abstract{We investigate gravitational capture of magnetic monopoles by primordial black holes (PBH) that evaporate before
Big Bang Nucleosynthesis (BBN), a hypothetical process which was once proposed as an alternative solution to the monopole
problem. Magnetic monopoles produced in phase transitions of a grand or partially unified gauge theory are considered. We prove analytically that for all extended PBH mass functions that preserve radiation domination, it is impossible to reduce the monopole abundance via gravitational capture by PBHs to values significantly below the one set by monopole annihilation
(or below its initial abundance if it is smaller),
regardless of the nature of the capture process (diffusive or non-diffusive). Therefore, the monopole problem cannot be solved
by PBH capture in a radiation-dominated era in the early universe.}

\begin{document}

\maketitle

\section{Introduction}
\label{sec:Intro}

Magnetic monopoles~\cite{Dirac:1931kp,tHooft:1974kcl,Polyakov:1974ek} arise as a class of topological objects in grand or partially unified theories with intriguing theoretical properties and important observational implications. Their magnetic charges are constrained by the Dirac quantization condition while their masses are tied to the corresponding unification scale. The abundance of magnetic monopoles is a cosmological issue~\cite{Preskill:1984gd}, which is one of the main driving forces behind the development of the inflation theory. Nevertheless, despite the great success of the inflation theory in solving the horizon and
flatness problems and providing a natural mechanism for generating the primordial density fluctuations that lead to structure formation, there have been a persistent interest in finding alternative solution to the monopole abundance problem, or more generally, the problem of overabundance of cosmological relics~\cite{Langacker:1980kd,Vilenkin:1981zs,Sikivie:1982qv,
Lazarides:1982tw,Izawa:1984ww,Salomonson:1984rh,Gelmini:1988sf,Dvali:1995cj,Bajc:1997ky,Dvali:1997sa,Stojkovic:2004hz,
Stojkovic:2005zh,Barr:2014vva,Kawasaki:2015lpf,Sato:2018nqy,Chatterjee:2019rch,Li:2023yzq}.

In the context of the monopole problem, whether magnetic monopoles are inflated away depends on whether they are produced before, during or after inflation. The current constraint on the scalar-to-tensor ratio~\cite{BICEP2:2015xme,Planck:2018vyg} sets an upper bound on the reheating temperature with the implication that magnetic monopoles associated with grand unification are likely to be diluted by inflation~\cite{ParticleDataGroup:2022pth,Lozanov:2019jxc}. Nevertheless, lighter monopoles, such as those associated with the Pati-Salam partially unified theories~\cite{Pati:1974yy,DiLuzio:2020xgc}, may also easily overclose the universe if they
are copiously produced. It is not clear whether such monopole problems are also solved by inflation, since the inflation energy scale is still unknown.

One alternative solution to the monopole problem is gravitational capture by primordial black holes (PBH) that evaporate before
Big Bang Nucleosynthesis (BBN)~\cite{Stojkovic:2004hz}. The implication of this solution goes beyond the monopole problem itself, as it would imply PBHs in the early universe may also significantly affect the relic abundance of other stable massive particles (SMP)~\cite{Burdin:2014xma}. Such SMPs may even serve as dark matter candidates. For example, in a dark sector in which a semi-simple dark gauge group is broken to a $U(1)$ subgroup, hidden monopoles arise as a result of the dark
gauge symmetry breaking and may also serve as dark matter candidate~\cite{Murayama:2009nj,GomezSanchez:2011orv,Evslin:2012fe,
Baek:2013dwa,Khoze:2014woa,Kawasaki:2015lpf,Nomura:2015xil,Sato:2018nqy,Daido:2019tbm,Bai:2020ttp,Graesser:2020hiv,
Nakagawa:2021nme,Graesser:2021vkr,Fan:2021ntg,Hiramatsu:2021kvu,Yang:2022quy}. It is thus motivated to consider the effect of PBHs on the abundance of such SMPs as well.

In a recent work ~\cite{Zhang:2023tfv} the present authors revisited the gravitational capture of magnetic monopoles by PBHs. It was found that the earlier analysis of ref.~\cite{Stojkovic:2004hz} overestimates the capture cross section in
the diffusive regime and thus leads to an overly optimistic assessment of the PBH capture capability. The point is that ref.~\cite{Stojkovic:2004hz} has modelled the PBH capture somewhat different from monopole annihilation while in fact these two
processes are quite similar. In ~\cite{Zhang:2023tfv} we have put the two processes on the same footing and explained the rationale
behind our treatment. For a monochromatic PBH mass function that preserves radiation domination it was shown that the PBH capture rate is several orders of magnitude below what is expected from the earlier treatment, and also far from what is needed to reduce the monopole abundance significantly. On the other hand, the earlier work ~\cite{Stojkovic:2004hz} actually proposed an extended
PBH mass function so that lighter PBHs keep evaporating while heavier PBHs keep forming to maintain radiation domination. Such an
extended PBH mass function is expected to lead to a better capture capability in the radiation-dominated era as it makes full use
of the energy density fraction that can be allocated to PBHs. It is therefore well-motivated to generalize our previous analysis to the case of such an extended PBH mass function as well, which is important for a proper assessment of the PBH capture capability regarding the relic abundance of various SMPs.

Such a generalization to the case of extended PBH mass functions while still assuming radiation domination is the main focus of the
present work. Assuming the PBHs account for a nearly fixed fraction of total energy density that is smaller than that of radiation,
it is possible to derive a functional equation for the PBH mass function. Nevertheless, it is not straightforward to analytically or numerically employ such a mass function if one only knows the functional equation it obeys. Here we utilize a special feature of
the diffusive PBH capture rate which can be easily generalized to the extended case to facilitate the analysis. In fact we introduce a parametrization of the capture term that is appropriate for both diffusive and non-diffusive capture. The special feature of diffusive capture allows us to prove easily that assuming radiation domination, the monopole abundance cannot
be reduced significantly by PBH capture (barring the effect of monopole annihilation) even if an extended PBH mass function is
allowed. Moreover, we carry over the analysis to the case of non-diffusive PBH capture. In the non-diffusive regime, the special
feature associated with the diffusive regime does not exist. Nevertheless, we prove via tricks of inequalities that assuming radiation domination the non-diffusive PBH capture cannot significantly reduce the monopole abundance either. Therefore, we reach a
quite generic conclusion that in the radiation-dominated era, the abundance of magnetic monopoles cannot be reduced significantly
by PBH capture to values below the one set by monopole annihilation (or below its initial abundance if it is smaller). Since it is well-known that monopole annihilation alone cannot solve the monopole problem~\cite{Weinberg:2012pjx,Goldman:1980sn}, this implies that PBH capture in a radiation-dominated era cannot solve the monopole problem either.

This work is organized as follows. In Sec.~\ref{sec:mips} we review the modelling of PBH capture of magnetic monopoles, explain the subtleties involved, and work out the various constraints on the parameter space from the assumption of existence of diffusive or
non-diffusive PBH capture of magnetic monopoles in the early universe. In Sec.~\ref{sec:dnd} we introduce a general parametrization
of the capture term, which is then employed to study both the diffusive and non-diffusive PBH capture, for both monochromatic and extended PBH mass functions assuming radiation domination. We prove the inability for PBH capture to significantly reduce the monopole abundance (barring the monopole annihilatiopn effect) in a radiation-dominated era. Finally we give our discussion and conclusions in Sec.~\ref{sec:dnc}.

\section{Modelling of PBH capture and parameter space}
\label{sec:mips}

\subsection{Monopole production and annihilation}

For definiteness we consider magnetic monopoles associated with a partially unified gauge theory such as the Pati-Salam model,
though the exact origin (i.e. the particle physics model behind) of magnetic monopoles is not essential to the analysis.
The virtue to consider partially unified gauge theories rather than grand unified theories is that their unification scales
are less constrained and may span a large range of energy scales depending on model construction. For example, the Pati-Salam breaking scale can range from just below the Planck scale, to as low as $\mO(10\TeV)$ depending on the field content~\cite{Jeannerot:2000sv,Hartmann:2014fya,DiLuzio:2020xgc,Dolan:2020doe,Cacciapaglia:2019dsq,Cacciapaglia:2020jvj}. Thus, the corresponding
magnetic monopole mass $m$ is also flexible. It is related to the symmetry breaking temperature $T_c$ via
\begin{align}
m=\delta T_c.
\label{eqn:deltax}
\end{align}
Typically $\delta=\mO(10)$. We will adopt the reference point value
\begin{align}
\delta=50,
\end{align}
The magnetic charge of the monopole is subject to the Dirac quantization condition. We parameterize the magnitude of its magnetic
charge as $\chi g$, in which
\begin{align}
g\simeq 5.9
\label{eqn:unitmag}
\end{align}
is the unit magnetic charge in natural Gaussian units, and $\chi$ is a positive integer determined by the particle physics model.
For example, Pati-Salam extensions of the Standard Model feature $\chi=2$ which we adopt as the reference value in this work, while trinification scenarios feature $\chi=3$~\cite{Lazarides:1986rt,Lazarides:1988wz,Kephart:2006zd,Kephart:2017esj,Lazarides:2021tua}.

The initial abundance of magnetic monopoles depends on the nature of the symmetry breaking phase transition that produces them. We assume radiation domination in the early universe, with the energy density $\rho$ and entropy density $s$ at temperature $T$ given by
\begin{align}
\rho=K_1 T^4,\quad s=K_2 T^3,
\label{eqn:rhos}
\end{align}
in which
\begin{align}
K_1=\frac{\pi^2}{30}\mN,\quad K_2=\frac{2\pi^2}{45}\mN,
\label{eqn:k1k2}
\end{align}
with $\mN$ being the number of effective relativistic degrees of freedom at temperature $T$. The Hubble parameter is then
\begin{align}
H=K\frac{T^2}{\MPl},
\label{eqn:hubble}
\end{align}
with
\begin{align}
K=\bigg(\frac{4\pi^3\mN}{45}\bigg)^{1/2},
\label{eqn:k}
\end{align}
and the Planck mass
\begin{align}
\MPl=1.2\times 10^{19}\GeV=2.2\times 10^{-5} \text{g}.
\label{eqn:Planckmass}
\end{align}
Causality considerations limit the maximum correlation length at a given temperature, which implies a lower bound
on the density of topological defects produced by the phase transition if it is associated with a nontrivial homotopy group~\cite{Kibble:1976sj}.
In the case of magnetic monopoles whose number density is denoted $n_M$, we introduce the monopole yield
\begin{align}
r\equiv\frac{n_M}{s},
\end{align}
and the reduced phase transition temperature
\begin{align}
x\equiv\frac{T_c}{\MPl},
\end{align}
then the lower bound on the initial monopole yield (Kibble estimate) can be expressed as~\cite{Zhang:2023tfv}
\begin{align}
r_i\gtrsim r_i^\text{Kibble},\quad r_i^\text{Kibble}=p(8\pi)^{3/2}\mN^{1/2}x^3,
\end{align}
where $p$ is a number not much less than $0.1$.

First-order phase transitions proceed via bubble nucleation. In such a case the initial monopole yield
can be expressed in terms of the bubble wall velocity $v_w$ and the $\beta$ parameter that characterizes
the inverse duration of the phase transition~\cite{Hindmarsh:2020hop}. More specifically we introduce
the dimensionless $\beta$ parameter
\begin{align}
\tilde\beta\equiv\frac{\beta}{H(T_p)},
\end{align}
with $T_p$ being the percolation temperature. Then the initial monopole yield $r_i$ can be expressed as
\begin{align}
r_i\simeq p(\tilde\beta v_w^{-1})^3 (8\pi)^{1/2}\mN^{1/2}x^3.
\label{eqn:ri1st}
\end{align}
Here $p$ is a number not much less than $0.1$. For a strongly first-order phase transition $\tilde\beta v_w^{-1}\simeq\mathcal{O}(1)$, the initial yield is close to the Kibble estimate\footnote{The Kibble estimate
does not apply to the case of extreme supercooling where $T_p/T_c\ll 1$, which we do not consider in this work.}, while for a typical weakly first-order phase transition $\tilde\beta v_w^{-1}\simeq\mathcal{O}(10\sim10^3)$, the initial yield can be orders of magnitude larger than the Kibble estimate.

For second-order phase transitions, the initial monopole yield is determined by the Kibble-Zurek mechanism~\cite{Zurek:1985qw,delCampo:2013nla}. The essential idea is the correlation length should be frozen
when the temporal distance to the critical point is equal to the equilibrium relaxation time. If we introduce
two critical exponents $\nu,\nu$ associated with the equilibrium correlation length and equilibrium relaxation time respectively
\begin{align}
\xi(\epsilon)=\frac{\xi_0}{|\epsilon|^\nu},\quad\tau(\epsilon)=\frac{\tau_0}{|\epsilon|^{\mu}},
\label{eqn:nudef}
\end{align}
the initial monopole yield can be expressed in terms of $\nu,\mu$ as
\begin{align}
r_i\simeq\lambda^{3/2}K_2^{-1}\bigg(\frac{Kx}{\sqrt{\lambda}}\bigg)^{\frac{3\nu}{1+\mu}},
\label{eqn:ri2ndb}
\end{align}
in which $\lambda\sim\mO(1)$ is a typical scalar quartic coupling in the theory. We typically consider
\begin{align}
\mu=\nu,\quad \lambda\simeq 1,
\end{align}
then
\begin{align}
r_i\simeq 0.02\bigg(\frac{17T_c}{\MPl}\bigg)^\frac{3\nu}{1+\nu},\quad\text{for}\,\,\mN=100,~\lambda=1.
\label{eqn:KZ}
\end{align}
Typically $\nu=0.5\sim 0.8$~\cite{Murayama:2009nj} and the resulting monopole yield is much larger than
that of the Kibble estimate.

After being produced in the phase transition, monopoles behave as nonrelativistic objects that are in
kinetic (but not chemical) equilibrium with the primordial plasma. The number of magnetic monopoles
in a comoving volume changes as a result of monopole annihilation~\cite{Zeldovich:1978wj,Preskill:1979zi}. The annihilation is mostly
in the diffusive regime which is characterized by two main features:
\begin{enumerate}
\item Monopoles and antimonopoles drift toward each other as a result of the balancing between the magnetic
attraction force and the drag force exerted by the plasma.
\item Monopoles and antimonopoles undergo Brownian motion in the plasma with a characteristic root-mean-square
velocity and mean free path.
\end{enumerate}
The drag force exerted on a nonrelativistic magnetic monopole can be approximated as~\cite{Vilenkin:2000jqa}
\begin{align}
\textbf{F}_\text{drag}=-CT^2\textbf{v},
\label{eqn:dragforce}
\end{align}
where $C\sim(1-5)\mN_c\chi^2$ with $\mN_c$ being the number of relativistic effective charged degrees of freedom~\cite{Vilenkin:2000jqa}. Therefore when the distance between monopole and antimonopole is $R$, the drift
velocity of the monopole is given by
\begin{align}
v_D\simeq\frac{\chi^2 g^2}{CT^2 R^2}.
\end{align}
Equating the negative Coulomb magnetic energy with the thermal kinetic energy of the monopole yields the
annihilation capture radius
\begin{align}
r_c=\frac{\chi^2 g^2}{T}.
\end{align}
The mean free path of the monopole is~\cite{Vilenkin:2000jqa,Weinberg:2012pjx}
\begin{align}
\ell\simeq\frac{1}{CT}\bigg(\frac{m}{T}\bigg)^{1/2}.
\label{eqn:mfp}
\end{align}
The diffusive capture is effective only if $r_c>\ell$, which translates into a condition on the temperature
\begin{align}
T>T_\text{ann},\quad T_\ann=\frac{m}{C^2\chi^4 g^4}.
\label{eqn:Tann}
\end{align}
In terms of the reduced temperature $z_\ann\equiv\frac{T_\ann}{T_c}$
\begin{align}
z_\ann=\delta C^{-2}\chi^{-4}g^{-4}.
\end{align}
The rationale behind the $r_c>\ell$ criterion is that once the distance between a monopole and an antimonopole
is smaller than $r_c$, then $r_c>\ell$ implies that thermal Brownian motion of the monopole (and antimonopole)
is unlikely to increase their distance to be larger than $r_c$ again. The motion of the pair will then be dominated
by the drifting so they are doomed to annihilate. Once the temperature drops below $T_\ann$, then $r_c<\ell$,
which implies with a distance between monopole and anitimonopole smaller than $r_c$ it is not guaranteed that the pair
will annihilate. In this non-diffusive capture regime the monopole and antimonopole lose their energy by
radiative capture via bremsstrahlung emission, and the corresponding capture rate turns out to be much smaller than
the diffusive capture regime~\cite{Weinberg:2012pjx}.

Taking into account of monopole annihilation, the monopole yield at $T=T_\ann$ can be estimated as follows.
The evolution of $n_M$ obeys the equation
\begin{align}
\dot{n}_M=-Dn_M^2-3\frac{\dot{a}}{a}n_M.
\label{eqn:nmevolve}
\end{align}
The $-3\frac{\dot{a}}{a}n_M$ term obviously takes into account of the effect of cosmic expansion, and the
annihilation capture coefficient $D$ can be computed from the characteristic capture time $\tau_\ann$
(with typical monopole separation $d_\text{ann}\sim n_M^{-1/3}$)
\begin{align}
\tau_\text{ann}\simeq \frac{d_\text{ann}}{v_\text{D}(d_\text{ann})}=\frac{CT^2}{\chi^2 g^2 n_M},
\end{align}
and~\cite{Weinberg:2012pjx}
\begin{align}
D=\frac{1}{\tau_\text{ann} n_M}=\frac{\chi^2 g^2}{CT^2}.
\label{eqn:D}
\end{align}
The evolution equation Eq.~\eqref{eqn:nmevolve} can then be solved analytically, and the monopole yield
$r_\ann$ at $T=T_\ann$ is found to be~\cite{Zhang:2023tfv}
\begin{align}
r_\ann\simeq\min\{r_i,r_\star\},\quad r_\star\equiv K_2^{-1}KC^{-1}\chi^{-6}g^{-6}\delta x.
\end{align}
Here $r_i$ is the initial monopole yield at $T=T_c$.

\subsection{Modelling of diffusive monopole capture by PBHs}

PBHs may form from primordial density fluctuations and a number of other mechanisms in the early universe~\cite{Sasaki:2018dmp,Khlopov:2008qy,Calmet:2015fua,Villanueva-Domingo:2021spv,Carr:2020gox,
Carr:2020xqk,Carr:2021bzv,Liu:2021svg,Escriva:2022duf}. Here we review gravitational capture of
magnetic monopoles by PBHs~\cite{Zhang:2023tfv}, which depends on the PBH mass and energy density fraction, but not its
formation mechanism. If the PBH mass function is monochromatic, then the PBHs are characterized by a single
mass $m_\bh$, or the reduced PBH mass parameter
\begin{align}
y\equiv\frac{m_\bh}{\MPl}.
\end{align}
The gravitational capture is similar to monopole annihilation in the sense that both are driven by long-range
forces that obey an inverse-square law. A gravitational capture radius which is the counterpart of $r_c$ in the
annihilation case can be likewise defined
\begin{align}
r_c^\gc=\frac{mm_\bh}{\MPl^2 T},
\label{eqn:rcgc}
\end{align}
The diffusive gravitational capture regime is characterized by $r_c^\gc>\ell$, which translates into a requirement
on the temperature
\begin{align}
T>T_\gc,\quad T_\gc\equiv\frac{\MPl^4}{C^2 m_\bh^2 m}.
\label{eqn:Tgc}
\end{align}
The corresponding reduced temperature $z_\gc$ is then
\begin{align}
z_\gc=C^{-2}\delta^{-1}x^{-2}y^{-2}.
\end{align}

The evolution of $n_M$ can be written as
\begin{align}
\dot{n}_M=-Dn_M^2-Fn_M-3\frac{\dot{a}}{a}n_M,
\end{align}
where the new term $-Fn_M$ embodies the effect of PBH capture. The coefficient $F$ can be
found as follows. The monopole drift velocity $u_D$ is a function of the monopole-PBH distance $R$
\begin{align}
u_\text{D}(R)=\frac{m_\bh m}{\MPl^2}\frac{1}{CT^2R^2}.
\end{align}
Suppose the number density of PBHs is $n_\bh$. The typical PBH separation is then $n_\bh^{-1/3}$, thus we use the typical drift velocity
\begin{align}
u_\text{D}(n_\bh^{-1/3})=\frac{m_\bh m}{\MPl^2}\frac{n_\bh^{2/3}}{CT^2}.
\end{align}
The typical gravitational capture time is
\begin{align}
\tau_\gc=\frac{n_\bh^{-1/3}}{u_\text{D}(n_\bh^{-1/3})}=\frac{\MPl^2 CT^2}{n_\bh m_\bh m}.
\label{eqn:captime}
\end{align}
$F$ should be interpreted as the typical capture frequency per monopole, and is given by
\begin{align}
F\equiv\tau_\gc^{-1}=\frac{n_\bh m_\bh m}{\MPl^2 CT^2}.
\label{eqn:fe1}
\end{align}
Alternatively, we may derive $F$ in the flux description~\cite{Zhang:2023tfv}. In the diffusive
regime, each PBH can be viewed as carrying a capture cross section
\begin{align}
\sigma_{g\text{D}}\equiv\pi (r_c^\gc)^2,
\label{eqn:sigmagD}
\end{align}
and being hit by monopoles with a characteristic incident velocity $v_{M\text{D}}$. The appropriate
choice for $v_{M\text{D}}$ is the drift velocity at a monopole-PBH distance $R=r_c^\gc$
\begin{align}
v_{M\text{D}}\equiv u_\text{D}(r_c^\gc)=\frac{m_\bh m}{\MPl^2}\frac{1}{CT^2(r_c^\gc)^2}=\frac{\MPl^2}{Cmm_\bh}.
\label{eqn:incident}
\end{align}
Therefore in the flux description, $F$ is found to be
\begin{align}
F=\sigma_{g\text{D}}v_{M\text{D}}n_\bh=\pi \frac{n_\bh m_\bh m}{\MPl^2 CT^2},
\label{eqn:Fpi}
\end{align}
which agrees with Eq.~\eqref{eqn:fe1} up to an $\mO(1)$ factor.

In the case of a monochromatic PBH mass function, the above expressions (Eq.~\eqref{eqn:fe1} and ~\eqref{eqn:Fpi}) for $F$ do not agree with ref.~\cite{Stojkovic:2004hz} which first proposed PBH capture as a solution to the monopole problem. The modelling of
ref.~\cite{Stojkovic:2004hz} would lead to $F\propto T^3$ while Eq.~\eqref{eqn:Fpi} leads to $F\propto T$ (assuming $n_\bh\propto T^3$). The two ways of modelling are compared in our previous work ~\cite{Zhang:2023tfv} which found that Eq.~\eqref{eqn:Fpi} leads
to a gravitational capture rate that is several orders of magnitude smaller than what is expected from ref.~\cite{Stojkovic:2004hz}. The difference originates from the use of the gravitational capture cross section: When both ways of modelling are framed in the flux language, it is found that the same monopole incident velocity is used, but the gravitational capture cross sections used in the two approaches are drastically different. Ref.~\cite{Stojkovic:2004hz} has used a gravitational
capture cross section that is derived from solving the geodesic motion of a test massive particle in a Schwarzschild geometry, which does not make sense in the diffusive regime when $r_c^\gc>\ell$. The reasonable estimate for the gravitational capture cross section in the diffusive regime should be given by $\pi (r_c^\gc)^2$, based on the physical picture of a competition between
monopole drift and random walk. In fact one may use a similar flux description for monopole annihilation. With an annihilation capture cross section estimated as $\pi r_c^2$ it is possible to derive the expression for $D$ (Eq.~\eqref{eqn:D}) again.

\subsection{Constraints on parameter space}

A number of constraints have to be taken into account when we analyze gravitational capture of magnetic monopoles by PBHs.
To this end, besides $T_\ann$ and $T_\gc$, we consider two more characteristic temperatures $T_\ev$ and $T_b$, associated with
PBH evaporation and formation, respectively.

$T_\ev$ is defined to be the temperature of the primordial plasma at the time of PBH evaporation, for a given PBH mass.
Since the PBH lifetime can be estimated as~\cite{Hooper:2020otu,Gehrman:2022imk}
\begin{align}
\tau_\bh=\varepsilon\frac{m_\bh^3}{\MPl^4},\quad \varepsilon=\frac{10240\pi}{\mathcal{G}\langle g_{\star,H}\rangle},
\label{eqn:vedef}
\end{align}
with $\mathcal{G}\simeq 3.8$ is the grey body factor, and $\langle g_{\star,H}\rangle\simeq\mN$ depends
on the particle physics model, equating $\tau_\bh$ with the cosmic time in a radiation-dominated era $t=\frac{1}{2H}$
determines $T_\ev$, or one may use the reduced temperature $z_\ev\equiv\frac{T_\ev}{T_c}$
\begin{align}
z_\ev=(2\varepsilon K)^{-1/2}x^{-1}y^{-3/2}.
\end{align}

$T_b$ is defined to be the temperature of the primordial plasma at the time of PBH evaporation, for a given PBH mass.
Usually the mass of a PBH at formation is linked to the horizon mass~\cite{Sasaki:2018dmp}
\begin{align}
m_\bh=\frac{\gamma}{2}\MPl^2 H_\text{form}^{-1},\quad H_\text{form}^{-1}=K\frac{T_b^2}{\MPl},
\label{eqn:gammadef}
\end{align}
Typically $\gamma\simeq 0.2$~\cite{Carr:1975qj}. Therefore the reduced temperature $z_b\equiv\frac{T_b}{T_c}$ associated
with PBH formation is
\begin{align}
z_b=\Big(\frac{\gamma}{2K}\Big)^{1/2}x^{-1}y^{-1/2}.
\end{align}

We summarize the parameters that appear in the analysis in Table~\ref{tab:parameters}, along with their definition
and reference point values. The ``Floating range'' column indicates the range of parameters taking into account of
uncertainties and the need to consider alternative scenarios. The expressions for the four reduced characteristic
temperatures $z_\ann,z_\gc,z_\ev,z_b$ are summarized in Table~\ref{tab:ztable}, along with the corresponding expressions
at the reference point.

\begin{table*}[t!]
\begin{center}
\begin{tabular}{|c|c|c|c|}
\hline
Parameter & Definition & Reference point value & Floating range \\
\hline
$\MPl$ & Eq.~\eqref{eqn:Planckmass} & $1.2\times 10^{19}\GeV$ & NA \\
\hline
$\mN$ & Eq.~\eqref{eqn:k1k2} & $100$ & $100\lesssim\mN\lesssim 1000$ \\
\hline
$g$ & Eq.~\eqref{eqn:unitmag} & 5.9 & $1\lesssim g\lesssim 10$ \\
\hline
$\chi$ & Integer magnetic charge & 2 & $\chi=1,\,2\,\text{or}\,3$ \\
\hline
$\delta$ & Eq.~\eqref{eqn:deltax} & $50$ & $10\lesssim\delta\lesssim 100 $ \\
\hline
$C$ & Eq.~\eqref{eqn:dragforce} & $200$ & $100\lesssim C\lesssim 1000$ \\
\hline
$\gamma$ & Eq.~\eqref{eqn:gammadef} & $0.2$ & $0.01\lesssim\gamma\lesssim 1$ \\
\hline
$\varepsilon$ & Eq.~\eqref{eqn:vedef} & 100 & $10\lesssim\varepsilon\lesssim 300$ \\
\hline
\end{tabular}
\caption{\label{tab:parameters} Summary of the parameters that appear in the analysis, with their definitions, reference point values, and floating range.}
\end{center}
\end{table*}

\begin{table*}[t!]
\begin{center}
\begin{tabular}{|c|c|c|}
\hline
Reduced temperature & Analytic expression & Reference point expression  \\
\hline
$z_b$ & $\big(\frac{\gamma}{2K}\big)^{1/2}x^{-1}y^{-1/2}$ & $0.078x^{-1}y^{-1/2}$ \\
\hline
$z_\ann$ & $\delta C^{-2}\chi^{-4} g^{-4}$ & $6.4\times 10^{-8}$ \\
\hline
$z_\ev$ & $(2\varepsilon K)^{-1/2}x^{-1}y^{-3/2}$ & $0.017x^{-1}y^{-3/2}$ \\
\hline
$z_\gc$ & $C^{-2}\delta^{-1}x^{-2}y^{-2}$ & $5\times 10^{-7}x^{-2}y^{-2}$ \\
\hline
\end{tabular}
\caption{\label{tab:ztable} Summary of the reduced characteristic temperatures.}
\end{center}
\end{table*}

We may derive constraints on the PBH mass by requiring the PBH form after inflation ($T_b\lesssim T_\text{RH}^{\text{max}}\simeq 10^{16}\GeV$) and evaporate before BBN $T_\ev\gtrsim T_\text{BBN}\simeq 1\MeV$. The resulting constraint on $y$ can be expressed as
\begin{align}
\frac{\gamma}{2K}\bigg(\frac{\MPl}{T_\text{RH}^\text{max}}\bigg)^2\lesssim y\lesssim (2\varepsilon K)^{-1/3}\bigg(\frac{\MPl}{T_\text{BBN}}\bigg)^{2/3},
\label{eqn:ymassc}
\end{align}
which reads at the reference point
\begin{align}
8.7\times 10^3\lesssim y\lesssim 3.5\times 10^{13}.
\end{align}

In the case of a monochromatic PBH mass function, we may derive a simple condition for radiation domination. To this
end we introduce the parameter $\beta$ (not to be confused with the phase transition parameter $\beta$), which is the ratio between PBH energy density and radiation energy density at PBH formation
\begin{align}
(n_\bh m_\bh)|_{\text{form}}=\beta K_1(T_b) T_b^4.
\label{eqn:betadef}
\end{align}
Since PBH is nonrelativistic, $n_\bh m_\bh\propto a^{-3}\propto s=K_2 T^3$, we find that at the time PBH evaporation
\begin{align}
(n_\bh m_\bh)|_{\text{evap}}=\beta\frac{K_1(T_b)}{K_2(T_b)}K_2(T_\ev)T_\ev^3 T_b.
\end{align}
If we require $(n_\bh m_\bh)|_{\text{evap}}$ be smaller than the radiation energy density at $T_\ev$, we find
\begin{align}
\frac{(n_\bh m_\bh)|_{\text{evap}}}{K_1(T_\ev)T_\ev^4}=\beta\frac{K_1(T_b)}{K_1(T_\ev)}\frac{K_2(T_\ev)}{K_2(T_b)}
\frac{T_b}{T_\ev}\lesssim 1.
\label{eqn:byc}
\end{align}
If we neglect the difference in $\mN$ defined in terms of the energy density and entropy density, then
\begin{align}
\frac{K_1(T_b)}{K_1(T_\ev)}\frac{K_2(T_\ev)}{K_2(T_b)}=1,
\end{align}
and Eq.~\eqref{eqn:byc} becomes
\begin{align}
\beta\frac{T_b}{T_\ev}\lesssim 1.
\end{align}
Using Table~\ref{tab:ztable} this condition is turned into
\begin{align}
\beta y\lesssim (\gamma\varepsilon)^{-1/2},
\label{eqn:radiationdom}
\end{align}
which reads at the reference point
\begin{align}
\beta y\lesssim 0.22.
\end{align}
It is interesting that this condition is not sensitive to the change of $\mN$ with respect to temperature.

In the following we generally use $z\equiv\frac{T}{T_c}$ to represent the reduced temperature. Let us
define two reduced temperatures $z_s$ and $z_t$ as
\begin{align}
z_s=\max\{z_\ev,z_\gc\},\quad z_t=\min\{1,z_b\}.
\end{align}
Diffusive PBH capture can only start at $z_t$ and end at $z_s$, so the existence of diffusive PBH
capture requires
\begin{align}
z_s<z_t,
\end{align}
which encodes four conditions, that is $z_\ev<1,z_\ev<z_b,z_\gc<1,z_\gc<z_b$. It turns out
that when we take into account the constraints on PBH mass Eq.~\eqref{eqn:ymassc}, the most constraining
condition among them is $z_\gc<1$, which can be translated into a constraint on $y$ for a given
value of $x$
\begin{align}
y>C^{-1}\delta^{-1/2}x^{-1},
\end{align}
which reads at the reference point
\begin{align}
y>7.1\times 10^{-4}x^{-1}.
\end{align}

In the following we will also be interested in non-diffusive gravitational capture of monopoles
by PBHs. The non-diffusive gravitational capture is possible because for a test massive particle
with nonrelativistic incident velocity $v$ at infinity there is a capture cross section
$\sigma_\text{nr}\approx\frac{4\pi R_\bh^2}{v^2}$, with $R_\bh$ being the Schwarzschild
radius of the PBH. This capture cross section is derived from solving the geodesic motion
of the test particle in the Schwarzschild geometry~\cite{Frolov:2011bhp}. In order to have non-diffusive
gravitational capture of magnetic monopoles, two conditions must be satisfied. The first is $z_\ev<1$, which
is equivalent to
\begin{align}
y>(2\varepsilon K)^{-1/3}x^{-2/3},
\label{eqn:257}
\end{align}
which reads at the reference point
\begin{align}
y>6.7\times 10^{-2}x^{-2/3}.
\end{align}
The second is $z_\ev<z_\gc$, which is equivalent to
\begin{align}
y<(2\varepsilon K)C^{-4}\delta^{-2}x^{-2},
\end{align}
which reads at the reference point
\begin{align}
y<8.3\times 10^{-10}x^{-2}.
\end{align}

In considering the capture of monopoles by PBHs, we implicitly treat monopoles like classical point
particles. Depending on the value of parameters, this approximation may break down. First, a magnetic
monopole produced in a phase transition at temperature $T_c$ is an extended object with classical size
\begin{align}
r_\cl=\frac{2g}{\chi T_c}.
\end{align}
Requiring $R_\bh>r_\cl$ gives
\begin{align}
y>\chi^{-1}gx^{-1},
\end{align}
which reads at the reference point
\begin{align}
y>3x^{-1}.
\end{align}
This constraint due to the monopole classical size is more stringent than both $z_\gc<1$ in the case of
diffusive capture and $z_\ev<1$ in the case of non-diffusive capture. Moreover, when the temperature drops
so that the electroweak symmetry is broken, the classical size of the monopole will grow to $v_\text{EW}^{-1}\sim 246\GeV^{-1}$, which is certainly larger than the Schwarzschild radius of the PBH which satisfies Eq.~\eqref{eqn:ymassc}.
However, at this moment it is not clear what will occur if a fat magnetic monopole encounters a small PBH with $R_\bh<r_\cl$.
It is possible that the PBHs still act as anchoring sites that facilitate the annihilation of monopoles and antimonopoles, but
in any case we do not expect the corresponding capture rate to be significantly enhanced relative to the $R_\bh>r_\cl$ case.
Therefore, the main conclusions of this work are not affected no matter we include or exclude the classical size
constraint $R_\bh>r_\cl$.

Another factor that might affect the PBH capture of magnetic monopoles is the quantum mechanical uncertainty. When the monopoles
are in kinetic equilibrium with the primordial plasma at temperature $T$, it acquires a mean thermal de Broglie wavelenth
$\lambda_\text{TdB}\sim(mT)^{-1/2}$. When $R_\bh<\lambda_\text{TdB}$, which can be realized in some portion of the parameter
space, the PBH will not be able to capture the monopole efficiently as usual. However, $\lambda_\text{TdB}$ is only appropriate
for monopoles far from the PBH. For monopoles close to the PBH drifting at high speed due to gravitational attraction, their
momenta are significantly increased relative to the thermal value, which in turn decreases their de Broglie wavelength and makes
the gravitational capture possible.

\begin{figure}[t]
\centering
	\includegraphics[width=0.8\linewidth]{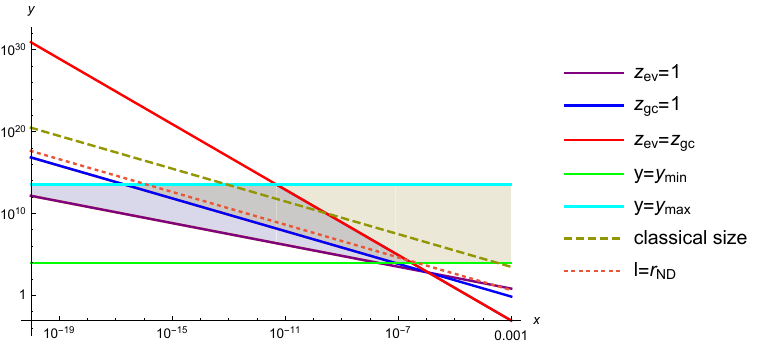}
	\caption{Illustration of the various lines of constraints and the region of diffusive and non-diffusive gravitational capture
on the $x-y$ plane (plotted on logarithmic scales). Parameters are taken using the reference point values in Table~\ref{tab:parameters}. The two horizontal solid lines correspond to the maximum and minimum PBH masses
allowed by Eq.~\eqref{eqn:ymassc}. The remaining solid lines correspond to $z_\ev=1$ (purple),$z_\gc=1$ (blue) and $z_\ev=z_\gc$ (red), respectively. The shadowed region to the right of the solid blue line allows to have diffusive gravitational capture, while
the shadowed region to the left of solid red line allows to have non-diffusive gravitational capture. The dark green dashed line corresponds to the classical size line $R_\bh=r_\cl$, while the red dashed line corresponds to the boundary when non-diffusive capture is saturated by the monopole mean free path for $z=1$.}
	\label{fig:diffusive}
\end{figure}

In Fig.~\ref{fig:diffusive} we plot the various lines of constraints and the region of diffusive and non-diffusive gravitational
capture on the $x-y$ plane, on logarithmic scales for both $x$ and $y$ (see figure caption for details). The range of $x$ under consideration is $10^{-20}\lesssim x\lesssim 10^{-3}$. If the classical size constraint $R_\bh>r_\cl$ is neglected, we see that diffusive gravitational capture is only possible for $x\gtrsim 2\times 10^{-17}$
while non-diffusive gravitational capture is only possible for $x\lesssim 3.1\times 10^{-7}$. In the overlapping shadowed region
both diffusive and non-diffusive capture are possible, corresponding to the case in which there is a transition from diffusive
to non-diffusive capture when the temperature drops below $T_\gc$.

\section{Diffusive and non-diffusive analyses of PBH capture with extended mass functions}
\label{sec:dnd}

\subsection{General evolution equations}

In all cases, the evolution of the monopole number density $n_M$ can be expressed as
\begin{align}
\dot{n}_M=-Dn_M^2-Fn_M-3\frac{\dot{a}}{a}n_M,
\label{eqn:gee}
\end{align}
where the $-Dn_M^2$ term characterizes the effect of monopole annihilation, the $-Fn_M$ term characterizes the effect
of PBH capture, and the $-3\frac{\dot{a}}{a}n_M$ characterizes the effect of cosmic expansion. Eq.~\eqref{eqn:gee}
is sufficiently general in that the feature of the annihilation or capture process can be encoded in the functional
form of the $D,F$ coefficients. With the introduction of the monopole yield $r\equiv\frac{n_M}{s}$, Eq.~\eqref{eqn:gee} can then be
transformed into
\begin{align}
\dot{r}=-Dsr^2-Fr,
\end{align}
which holds even if $\mN$ is a function of temperature. Next we make the time-to-temperature transition, which leads to
\begin{align}
\frac{dr}{dT}=\frac{Ds}{HT}r^2+\frac{F}{HT}r.
\label{eqn:drdT}
\end{align}
This equation holds when we ignore the temperature dependence of $\mN$, but the dependence of $D,F$ on temperature
is not restricted.

We now introduce the power-law parametrization of $D$ and $F$ which is applicable to a wide variety of scenarios:
\begin{align}
D=D_0 T^{n_D},\quad F=F_0 T^{n_F},
\end{align}
with $D_0,F_0$ being temperature-independent functions of $x,y$ carrying the appropriate mass dimensions, and $n_D,n_F$ are
real constants determined by the corresponding annihilation or capture scenarios. Let us define
\begin{align}
w=-\ln z,\quad\text{or}\quad z=\exp(-w).
\end{align}
Then Eq.~\eqref{eqn:drdT} can be turned into
\begin{align}
\frac{d\ln r}{dw}=-J_De^{-(n_D+1)w}r-J_Fe^{-(n_F-2)w},
\label{eqn:dlnrdw}
\end{align}
where
\begin{align}
J_D=D_0K_2K^{-1}\MPl T_c^{n_D+1},\quad J_F=F_0K^{-1}\MPl T_c^{n_F-2}.
\end{align}
When both $J_D,J_F$ terms are present, Eq.~\eqref{eqn:dlnrdw} does not allow for simple analytic solutions except for
some special cases.

For example, in the case of diffusive annihilation, $D$ is given by Eq.~\eqref{eqn:D}. This is equivalent to
\begin{align}
D_0=C^{-1}\chi^2 g^2,\quad n_D=-2,\quad\text{(Diffusive monopole annihilation)}
\end{align}
and $J_D$ is given by
\begin{align}
J_D=C^{-1}K^{-1}K_2\chi^2 g^2 x^{-1},\quad\text{(Diffusive monopole annihilation)}
\end{align}
If PBH capture can be neglected, the evolution equation Eq.~\eqref{eqn:dlnrdw} becomes
\begin{align}
\frac{d\ln r}{dw}=-J_De^wr,\quad\text{(Diffusive monopole annihilation)}.
\end{align}

If one considers non-diffusive annihilation, then $D$ is given by~\cite{Vilenkin:2000jqa,Weinberg:2012pjx,Zeldovich:1978wj,Dicus:1982ri}
\begin{align}
D=\Big(\frac{g^2}{m}\Big)^2\Big(\frac{m}{T}\Big)^{9/10},\quad\text{(Non-diffusive monopole annihilation)}
\end{align}
which amounts to
\begin{align}
D_0=g^4 m^{-11/10},\quad n_D=-\frac{9}{10},\quad\text{(Non-diffusive monopole annihilation)}.
\label{eqn:ndd0nd}
\end{align}
If PBH capture can be neglected, the evolution equation Eq.~\eqref{eqn:dlnrdw} becomes
\begin{align}
\frac{d\ln r}{dw}=-J_De^{-w/10}r,\quad\text{(Non-diffusive monopole annihilation)}
\end{align}
where $J_D$ is computed using Eq.~\eqref{eqn:ndd0nd}.

For diffusive capture with a monochromatic PBH mass function, which is the focus of ref.~\cite{Zhang:2023tfv},
$F$ is given by Eq.~\eqref{eqn:fe1}. One may trade $n_\bh$ for the $\beta$ parameter to express $F$ as
(neglecting the dependence of $\mN$ on temperature)
\begin{align}
F=\beta\delta K_1C^{-1}\Big(\frac{\gamma}{2K}\Big)^{1/2}xy^{-1/2}T,\quad\text{(Diffusive PBH capture,monochromatic)}
\end{align}
which amounts to
\begin{align}
F_0=\beta\delta K_1C^{-1}\Big(\frac{\gamma}{2K}\Big)^{1/2}xy^{-1/2},\quad n_F=1,\quad\text{(Diffusive PBH capture,monochromatic)}.
\end{align}
The corresponding $J_F$ is computed to be
\begin{align}
J_F=\Big(\frac{\gamma}{2}\Big)^{1/2}K_1 K^{-3/2}C^{-1}\delta\beta y^{-1/2},\quad\text{(Diffusive PBH capture,monochromatic)}.
\end{align}
This expression of $J_F$ corresponds to $\bar{\Phi}$ in ref.~\cite{Zhang:2023tfv}. If monopole annihilation can be neglected,
then
\begin{align}
\frac{d\ln r}{dw}=-J_F e^w,\quad\text{(Diffusive PBH capture,monochromatic)}.
\end{align}
If monopole annihilation is also in the diffusive regime, then
\begin{align}
&\frac{d\ln r}{dw}=-J_D e^w r-J_F e^w,\nonumber \\
&\text{(Diffusive annihilation and diffusive PBH capture,monochromatic)}.
\label{eqn:dlnrdwdadc}
\end{align}
Since the right-hand side of Eq.~\eqref{eqn:dlnrdwdadc} is proportional to $e^w$, it also allows for an analytic solution,
which is presented in ref.~\cite{Zhang:2023tfv}.

\subsection{Extended PBH mass function}

If we consider an extended PBH mass function, the expression for $F$ should be generalized from
the monochromatic case Eq.~\eqref{eqn:fe1} to
\begin{align}
F=\frac{m}{\MPl CT^2}\int n(y,z)yd\ln y,
\end{align}
where
\begin{align}
n(y,z)\equiv\frac{dn_\bh}{d\ln y}.
\end{align}
The physical meaning of $n(y,z)$ is: at the reduced temperature $z$, the PBHs with a reduced mass
associated with the logarithmic interval $[\ln y,\ln y+d\ln y]$ have number density $n(y,z)d\ln y$.

For a given value of $y$, there are the reduced characteristic temperatures $z_b,z_\ev$ associated with
PBH formation and evaporation, see Table~\ref{tab:ztable}. These relations can be inverted to find
the reduced mass $y_b$ of the PBHs that form at the reduced temperature $z$ and the reduced mass $y_\ev$
of the PBHs that just evaporate at the reduced temperature $z$
\begin{align}
y_b=\frac{\gamma}{2K}x^{-2}z^{-2},\quad y_\ev=(2\varepsilon K)^{-1/3}x^{-2/3}z^{-2/3}.
\end{align}
Then $n(y,z)$ should vanish if $y<y_\ev$ or $y<y_b$ at any given reduced temperature $z$. On the other hand,
for $y_\ev\leq y\leq y_b$, $n(y,z)$ should scale as $z^3$, which reflects the fact that the number density of nonrelativistic objects that are not created or destroyed in the early universe should scale as $T^3$. Therefore, generally we may
parameterize $n(y,z)$ as
\begin{align}
n(y,z)=A(y)T_c^3 z^3 \theta(y-y_\ev)\theta(y_b-y),
\end{align}
where $A(y)$ is a dimensionless function of $y$ only, and $\theta$ denotes the Heaviside step function.

In order to maximally employ the capture capability of PBHs while retaining the radiation domination condition, we may
envision a kind of PBH mass function (i.e. a choice of $A(y)$) such that the energy density of PBHs remain a constant
fraction $f\leq 1$ of the radiation energy density, due to the constant formation and evaporation of PBHs. Strictly speaking, such a requirement can only be satisfied
for an intermediate range of reduced temperature. This is because we require PBHs form after inflation and evaporate before
BBN. Thus at very high temperature there is a period when the PBH energy density fraction starts to grow and at temperatures close
to BBN there is a period when the PBH energy density fraction gradually drops to zero. Despite this complication, let us
focus our attention in the intermediate temperature range when the PBH energy density fraction is indeed a constant. Suppose the
PBH energy density is $\rho_\bh$ and the radiation energy density is $\rho_r$, by analyzing
the equation
\begin{align}
\frac{d(\rho_\bh/\rho_r)}{d\ln z}=0,
\end{align}
we may arrive at a functional equation that should be satisfied by $A(y)$. Introducing
\begin{align}
B(y)\equiv A(y)y,
\end{align}
and
\begin{align}
k_0=f K_1 x,
\end{align}
this functional equation can be expressed as
\begin{align}
6B(y_b(z))-2B(y_\ev(z))+3k_0z=0,
\label{eqn:feB}
\end{align}
which should be satisfied for all $z$ in the above-mentioned intermediate range of the reduced temperature. Moreover,
there is the normalization condition
\begin{align}
\int_{y_\ev(z)}^{y_b(z)}B(y)d\ln y=k_0 z,
\label{eqn:norB}
\end{align}
which should be satisfied by all $z$ when $f$ is held constant.

Although it is possible to find the functional equation Eq.~\eqref{eqn:feB} and the normalization condition
Eq.~\eqref{eqn:norB} that deliver the desired PBH mass function, it is not straightforward to subject them to
analytic or numerical analyses. The desired PBH mass function depends on the implementation at the high and low
temperature ends and might not be unique. In the following, instead of using some analytic or numerical implementation
of mass functions satisfying Eq.~\eqref{eqn:feB} and Eq.~\eqref{eqn:norB}, we will take advantage of important
features of the capture term in relevant cases to facilitate the analyses in this work.

\subsection{Diffusive gravitational capture in the extended case}

The key to analyzing diffusive gravitational capture for extended PBH mass function is the observation that
in the monochromatic case, according to Eq.~\eqref{eqn:fe1}, $F\propto n_\bh m_\bh$ at a given temperature, while
$n_\bh m_\bh$ is just the energy density of PBHs. Therefore, when we consider an extended PBH mass function, at a
given temperature we should have $F\propto\rho_\bh$. This implies that $F$ is only sensitive to the total
energy density of PBHs but not the differential PBH mass distribution. Assuming radiation domination, let us consider
\begin{align}
\rho_\bh=fK_1 T^4,
\end{align}
where $0\leq f\leq1$ is a constant. Assuming all the PBH energy density contributes to diffusive gravitational capture,
we obtain
\begin{align}
F=\frac{fK_1 T^4 m}{\MPl^2 CT^2},
\end{align}
which can be expressed as
\begin{align}
F=fK_1 C^{-1}\delta x\MPl^{-1}T^2.
\end{align}
This corresponds to
\begin{align}
F_0=fK_1 C^{-1}\delta x\MPl^{-1},\quad n_F=2.
\end{align}
The corresponding $J_F$ is
\begin{align}
J_F=fK_1 K^{-1}C^{-1}\delta x,
\end{align}
which reads
\begin{align}
J_F\simeq 0.5fx
\label{eqn:JFrf}
\end{align}
at the reference point.

Suppose the annihilation term can be neglected, which is the case for $r_i<r_\star$ or $z<z_\ann$. The evolution equation
can be cast into
\begin{align}
\frac{d\ln r}{dw}=-J_F.
\label{eqn:dlnrdw01}
\end{align}
At the reference point $J_F$ is given by Eq.~\eqref{eqn:JFrf}, which is much smaller than $1$ for the range of $x$ under consideration ($x\lesssim 10^{-3}$). This means that even with an extended PBH mass function, the fractional efficiency
$-\frac{d\ln r}{dw}$ of monopole abundance reduction by diffusive capture is always smaller than $1$ for the range of $x$ under consideration. However, we should also check the cumulative effect of diffusive capture since the cosmic time increases by many
orders of magnitude. To this end, we note that since $J_F$ is temperature-independent, Eq.~\eqref{eqn:dlnrdw01} can be analytically solved
\begin{align}
r_2=r_1\exp[-J_F(w_2-w_1)],
\end{align}
where the subscripts $1$ and $2$ refer to the corresponding quantity at two arbitrary temperatures $T_1$ and $T_2$ respectively
($T_1>T_2$). To have a significant reduction of the monopole abundance, one would need a sufficiently long duration of diffusive gravitational capture, that is, $J_F(w_2-w_1)\gg 1$. For the largest value of $x$ under consideration $x\simeq 10^{-3}$,
$w_2-w_1=\ln(z_1/z_2)$ reaches its maximum $w_2-w_1\simeq 41$, while $J_F$ also reaches its maximum, so that
\begin{align}
J_F(w_2-w_1)\simeq 0.02f\ll 1.
\end{align}
This proves that with an extended PBH mass function, it is not possible to reduce the monopole abundance significantly with diffusive gravitational capture, assuming radiation domination. The above derivation assumes $x\lesssim 10^{-3}$, but obviously
it is still valid as long as $x\lesssim 0.1$ (which is roughly equivalent to a sub-Planckian monopole mass).

\subsection{Non-diffusive gravitational capture in the monochromatic case}

We now turn to non-diffusive gravitational capture of magnetic monopoles by PBHs with a monochromatic PBH mass function.
This non-diffusive gravitational capture regime is possible if $z_\ev<z_\gc$ and $z_\ev<1$, corresponding to the shadowed
region to the left of the red solid line in Fig.~\ref{fig:diffusive}. The corresponding requirement on $x$ is $x\lesssim 3.1\times
10^{-7}$ at the reference point. If one further imposes the classical size constraint $R_\bh>r_\cl$ which excludes the region
below the dashed dark green line in Fig.~\ref{fig:diffusive}, the allowed parameter space for non-diffusive gravitational capture
would be the small triangle bounded by the solid cyan and red lines and the dashed dark green line in Fig.~\ref{fig:diffusive}.
Nevertheless, the conclusions of this work do not depend on whether we impose the classical size constraint, thus we will not be
restricted by it in the following.

The capture coefficient $F$ in the flux description should be given by
\begin{align}
F=n_\bh\sigma_\gND v_M,
\end{align}
where $v_M=(3T/m)^{1/2}$ is the thermal velocity of the monopole, and $\sigma_\gND$ is the effective capture cross section
in the non-diffusive regime, which is given by
\begin{align}
\sigma_\gND=\min\{\pi r_\ND^2,\pi\ell^2\}.
\label{eqn:sgND}
\end{align}
Here $\ell$ is the monopole mean free path (see Eq.~\eqref{eqn:mfp}), and $r_\ND$ is given by~\cite{Frolov:2011bhp}
\begin{align}
r_\ND\equiv\frac{2R_\bh}{v_M}.
\end{align}
As discussed above Eq.~\eqref{eqn:257}, $r_\ND$ is obtained by solving the geodesic motion of
a test nonrelativistic particle in the Schwarzschild geometry (with an incident velocity $v_M$).
In Eq.~\eqref{eqn:sgND} the effective capture cross section is determined by a comparison between $r_\ND$
and $\ell$, because in the non-diffusive regime characterized by $\ell>r_c^\gc$, the capture should be
limited by the monopole mean free path. The boundary $\ell=r_\ND$ when $z=1$ is shown as the red dashed line in Fig.~\ref{fig:diffusive}. Below this boundary, $\sigma_\gND$ is always given by $\pi r_\ND^2$.

In any case, $\pi\ell^2$ sets an upper limit of the capture cross section in the non-diffusive regime. Therefore,
let us simply replace $\sigma_\gND$ with
\begin{align}
\pi\ell^2=\frac{\pi m}{C^2 T^3},
\end{align}
which can only overestimate the capture rate. The corresponding expression for $F$ is
\begin{align}
F=\frac{\pi n_\bh\sqrt{3mT}}{C^2 T^3}.
\end{align}
Now we trade $n_\bh$ for the $\beta$ parameter introduced in Eq.~\eqref{eqn:betadef}, so that $F$ is expressed as
\begin{align}
F=\pi C^{-2}\beta K_1\Big(\frac{\gamma}{2K}\Big)^{1/2}y^{-3/2}(3m)^{1/2}T^{1/2},
\label{eqn:Fndm}
\end{align}
which corresponds to
\begin{align}
F_0=\pi C^{-2}\beta K_1\Big(\frac{\gamma}{2K}\Big)^{1/2}y^{-3/2}(3m)^{1/2},\quad n_F=\frac{1}{2}.
\end{align}
The expression for $J_F$ is
\begin{align}
J_F=\pi\Big(\frac{3\gamma}{2}\Big)^{1/2}K_1 K^{-3/2}C^{-2}\delta^{1/2}x^{-1}\beta y^{-3/2},
\end{align}
which reads at the reference point
\begin{align}
J_F=1.5\times 10^{-4}x^{-1}\beta y^{-3/2}.
\end{align}
Thus, if the monopole annihilation term can be neglected, Eq.~\eqref{eqn:dlnrdw} becomes
\begin{align}
\frac{d\ln r}{dw}=-J_Fe^{3w/2}.
\label{eqn:dlnrdwndm}
\end{align}
Strictly speaking, this applies to the case in which $z\geq\frac{\sqrt{3}}{2}C^{-1}x^{-1}y^{-1}$. For more general
cases, Eq.~\eqref{eqn:dlnrdwndm} overestimates the non-diffusive gravitational capture rate.

Eq.~\eqref{eqn:dlnrdwndm} can be analytically solved:
\begin{align}
r_2=r_1\exp\Bigg[-\frac{2}{3}J_F(z_2^{-3/2}-z_1^{-3/2})\Bigg].
\end{align}
For non-diffusive capture let us consider $z_2=z_\ev$, we find that
\begin{align}
\frac{2}{3}J_Fz_\ev^{-3/2}=\frac{2}{3}\pi\Big(\frac{3\gamma}{2}\Big)^{1/2}(2\varepsilon K)^{-3/4}
K_1 K^{-3/2}C^{-2}\delta^{1/2}x^{1/2}\beta y^{3/4}.
\end{align}
Assuming the radiation domination condition Eq.~\eqref{eqn:radiationdom} is saturated, then
\begin{align}
\frac{2}{3}J_Fz_\ev^{-3/2}=2^{-1/4}3^{-1/2}\pi\varepsilon^{-5/4}K_1K^{-9/4}C^{-2}\delta^{1/2}x^{1/2}y^{-1/4},
\end{align}
which reads at the reference point
\begin{align}
\frac{2}{3}J_Fz_\ev^{-3/2}\simeq 1.4\times 10^{-4}x^{1/2}y^{-1/4}.
\end{align}
For the range of $x,y$ that allows for non-diffusive gravitational capture, we have
\begin{align}
\frac{2}{3}J_F(z_\ev^{-3/2}-z_1^{-3/2})<\frac{2}{3}J_Fz_\ev^{-3/2}\simeq 1.4\times 10^{-4}x^{1/2}y^{-1/4}\ll 1.
\end{align}
Thus for non-diffusive gravitational capture, it is not possible to achieve $r(z_\ev)\ll r_1$ for a monochromatic
PBH mass function assuming radiation domination. This conclusion is robust against possible variation of the
parameters according to the ``Floating range'' listed in Table~\ref{tab:parameters}.

\subsection{Non-diffusive gravitational capture in the extended case}

We now consider non-diffusive gravitational capture of magnetic monopoles by PBHs, assuming an extended PBH
mass function that preserves the radiation domination condition. This is more complicated than the corresponding
diffusive case as the capture coefficient $F$ is not simply proportional to $\rho_\bh$ at a given temperature.
Instead, $F$ is generalized from Eq.~\eqref{eqn:Fndm} to
\begin{align}
F=\pi C^{-2}K_1\Big(\frac{\gamma}{2K}\Big)^{1/2}(3m)^{1/2}T^{1/2}\sum_i\beta_iy_i^{-3/2}\theta(z-z_{\ev i})\theta(z_{bi}-z).
\label{eqn:Fnde}
\end{align}
Here we have divided the PBH mass range into a sufficiently large number of small bins, with the $i$th bin characterized by
its reduced mass $y_i$ and $\beta$ parameter $\beta_i$. $\beta_i$ is defined via
\begin{align}
(n_{\bh i} m_{\bh i})|_{\text{form}}=\beta_i K_1(T_{bi}) T_{bi}^4.
\label{eqn:betaidef}
\end{align}
$m_{\bh i}$ and $n_{\bh i}$ are the PBH mass and number density associated with the $i$th bin, respectively. The summation in Eq.~\eqref{eqn:Fnde} is over all PBH mass bins.

In order to proceed, we note that the radiation domination constraint can be expressed in the extended case as
\begin{align}
\sum_i n_{\bh i} m_{\bh i}\theta(z-z_{\ev i})\theta(z_{bi}-z)\lesssim K_1 T^4.
\label{eqn:rdce}
\end{align}
If we neglect the temperature dependence of $\mN$, then Eq.~\eqref{eqn:rdce} becomes
\begin{align}
\sum_i\beta_i z_{bi}\theta(z-z_{\ev i})\theta(z_{bi}-z)\lesssim z,\quad\text{at any }z.
\end{align}
Using the expression for $z_b$ in Table~\ref{tab:ztable}, we obtain
\begin{align}
\sum_i\beta_i y_i^{-1/2}\theta(z-z_{\ev i})\theta(z_{bi}-z)\lesssim \Big(\frac{\gamma}{2K}\Big)^{-1/2}xz,\quad\text{at any }z.
\label{eqn:ineq01}
\end{align}
Note the similarity between the summation in Eq.~\eqref{eqn:Fnde} and Eq.~\eqref{eqn:Fnde}: the only difference
is the power on $y_i$. This suggests using the trick of expanding or shrinking. If the reduced PBH mass is bounded from below
for all mass bins under consideration
\begin{align}
y_i\geq y_m,\quad \forall i,
\end{align}
we may expand the summation in Eq.~\eqref{eqn:Fnde} as
\begin{align}
\sum_i\beta_iy_i^{-3/2}\theta(z-z_{\ev i})\theta(z_{bi}-z)\leq y_m^{-1}\sum_i\beta_iy_i^{-1/2}\theta(z-z_{\ev i})\theta(z_{bi}-z).
\end{align}
Therefore, we may use Eq.~\eqref{eqn:ineq01} to obtain
\begin{align}
\sum_i\beta_iy_i^{-3/2}\theta(z-z_{\ev i})\theta(z_{bi}-z)\leq \Big(\frac{\gamma}{2K}\Big)^{-1/2}xzy_m^{-1},\quad\text{at any }z.
\end{align}
Then at all temperature $F$ from non-diffusive capture is bounded by
\begin{align}
F\leq F_m\equiv\pi C^{-2}K_1(3m)^{1/2}\MPl^{-1}y_m^{-1}T^{3/2}.
\end{align}
In the following let us simply consider $F=F_m$, which necessarily overestimates the capture rate. This corresponds to
\begin{align}
F_0=\pi C^{-2}K_1(3m)^{1/2}\MPl^{-1}y_m^{-1},\quad n_F=\frac{3}{2}.
\end{align}
The corresponding $J_F$ is given by
\begin{align}
J_F=\sqrt{3}\pi\delta^{1/2}C^{-2}K^{-1}K_1y_m^{-1}.
\end{align}
If the monopole annihilation can be neglected, Eq.~\eqref{eqn:dlnrdw} becomes
\begin{align}
\frac{d\ln r}{dw}=-J_Fe^{w/2}.
\label{eqn:dlnrdwnde}
\end{align}
Eq.~\eqref{eqn:dlnrdwnde} can be solved analytically
\begin{align}
r_2=r_1\exp\Big[-2J_F(z_2^{-1/2}-z_1^{-1/2})\Big].
\label{eqn:red0}
\end{align}
With this solution we find that it is not quite helpful to consider a universal $y_m$ for
the range of parameters that may produce non-diffusive gravitational capture. The reason is simple
to understand: for the parameter range associated with non-diffusive gravitational capture shown
in Fig.~\ref{fig:diffusive}, in a large portion of region the actual value of $y$ is larger than
$y_m$ by many orders of magnitude. Thus using a universal $y_m$ worsens significantly the power
of the inequality. Nevertheless, this weakness is easy to fix. We may simply divide the evolution
of monopole yield into multiple stages, with each stage a corresponding value of $y_m$. For example,
let us consider three stages of evolution:
\begin{align}
&\text{Stage I:}\quad z=1\rightarrow z=z_{\text{mid}1}\equiv 10^{-4}, \nonumber \\
&\text{Stage II:}\quad z=z_{\text{mid}1}\rightarrow z=z_{\text{mid}2}\equiv 10^{-8}, \nonumber \\
&\text{Stage III:}\quad z=z_{\text{mid}2}\rightarrow z=z_{\text{min}}\equiv \frac{T_\text{BBN}}{T_c}.
\end{align}
For each of the three stages, we use a corresponding value of $y_m$, and the value of $J_F$ is determined accordingly.
To be explicit, let us introduce
\begin{align}
x_\text{max}\equiv 2\varepsilon^{1/2}\gamma^{-1/2}KC^{-2}\delta^{-1}\frac{T_\text{RH}^\text{max}}{\MPl}
\end{align}
which reads $x_\text{max}\simeq 3.1\times 10^{-7}$ at the reference point. $x_\text{max}$ is just the maximum
value of $x$ allowed for non-diffusive gravitational capture. We also introduce
\begin{align}
x_{\text{mid}1}\equiv z_{\text{mid}1}x_\text{max},\quad x_{\text{mid}2}\equiv z_{\text{mid}2}x_\text{max}.
\end{align}
Then $x_{\text{mid}1}\MPl$ corresponds to the maximum temperature in Stage II, while $x_{\text{mid}2}\MPl$
corresponds to the maximum temperature in Stage III. Then $y_m$ for three stages are determined as follows
\begin{align}
y_m=y_{m1}\equiv\frac{\gamma}{2K}\bigg(\frac{\MPl}{T_\text{RH}^\text{max}}\bigg)^2,&\quad\text{Stage I},\nonumber \\
y_m=y_{m2}\equiv(2\varepsilon K)^{-1/3}x_{\text{mid}1}^{-2/3},&\quad\text{Stage II},\nonumber \\
y_m=y_{m3}\equiv(2\varepsilon K)^{-1/3}x_{\text{mid}2}^{-2/3},&\quad\text{Stage III}.
\label{eqn:ym123}
\end{align}
The choice of $y_m$ in three stages are motivated by the non-diffusive gravitational capture region in Fig.~\ref{fig:diffusive}.
For example, $y_{m1}$ comes from requiring the PBH be formed after inflation, $y_{m2}$ and $y_{m2}$ come from the
$z_\ev<1$ requirement in Fig.~\ref{fig:diffusive}, which translate into $z_\ev<z_{\text{mid}1}$ and $z_\ev<z_{\text{mid}2}$
in the current setting. At the reference point $y_{m2}\simeq 6.8\times 10^5,y_{m3}\simeq 3.2\times 10^8$. $J_F$'s in three
stages are then given by
\begin{align}
J_F=J_{F1}\equiv\sqrt{3}\pi\delta^{1/2}C^{-2}K^{-1}K_1y_{m1}^{-1},&\quad\text{Stage I}, \nonumber \\
J_F=J_{F2}\equiv\sqrt{3}\pi\delta^{1/2}C^{-2}K^{-1}K_1y_{m2}^{-1},&\quad\text{Stage II}, \nonumber \\
J_F=J_{F3}\equiv\sqrt{3}\pi\delta^{1/2}C^{-2}K^{-1}K_1y_{m3}^{-1},&\quad\text{Stage III}.
\end{align}
Eq.~\eqref{eqn:red0} is generalized to
\begin{align}
& r(z_{\text{mid}1})=r_i\exp\Big[-2J_{F1}(z_{\text{mid}1}^{-1/2}-1)\Big],\nonumber \\
& r(z_{\text{mid}2})=r(z_{\text{mid}1})\exp\Big[-2J_{F2}(z_{\text{mid}2}^{-1/2}-z_{\text{mid}1}^{-1/2})\Big],\nonumber \\
& r(z_\text{min})=r(z_{\text{mid}2})\exp\Big[-2J_{F3}(z_\text{min}^{-1/2}-z_{\text{mid}2}^{-1/2})\Big].
\end{align}
We therefore see that the effect of non-diffusive gravitational capture in each stage can be characterized
by the corresponding exponent (barring the negative sign), which we call the \emph{reduction exponent}. For example, the reduction exponent
associated with Stage I is $2J_{F1}(z_{\text{mid}1}^{-1/2}-1)$, and likewise for Stage II and III. The reduction
exponents of the three stages may be added cumulatively to characterize the total effect
of non-diffusive gravitational capture. A significant reduction of monopole yield is possible only if
the total reduction exponent is much larger than $1$. In the current setting where we only have three stages,
the reduction exponent of at least one stage must be much larger than $1$ to allow for a significant reduction
of the monopole yield. However, we now show that this is impossible assuming radiation domination. For definiteness, in Stage III
let us set
\begin{align}
z_\text{min}=\frac{T_\text{BBN}}{T_{c\text{max}}}=\frac{T_\text{BBN}/\MPl}{x_\text{max}},
\end{align}
which reads $z_\text{min}\simeq 2.7\times 10^{-16}$ at the reference point. This allows to maximize the capture effect
in Stage III. The reduction exponents in three stages, along with their values calculated at the reference point, are
(the reduction exponents are dominated by the term associated with the lower end of $z$, which we retain as a good approximation)
\begin{align}
2J_{F1}z_{\text{mid}1}^{-1/2}=4\sqrt{3}\pi\delta^{1/2}C^{-2}K_1\gamma^{-1}
\Big(\frac{T_\text{RH}^\text{max}}{\MPl}\Big)^2z_{\text{mid}1}^{-1/2}\simeq 4.4\times 10^{-5},\nonumber \\
2J_{F2}z_{\text{mid}2}^{-1/2}=2\sqrt{3}\pi\delta^{1/2}C^{-2}K^{-1}K_1y_{m2}^{-1}z_{\text{mid}2}^{-1/2}\simeq 5.6\times 10^{-5},\nonumber \\
2J_{F3}z_{\text{min}}^{-1/2}=2\sqrt{3}\pi\delta^{1/2}C^{-2}K^{-1}K_1y_{m3}^{-1}z_{\text{min}}^{-1/2}\simeq 7.4\times 10^{-4}.
\end{align}
We see that at the reference point, all reduction exponents are much smaller than $1$, and thus a significant reduction of
the monopole yield is not possible. Moreover, this conclusion is also robust against possible variation of the parameters
according to the ``Floating range'' listed in Table~\ref{tab:parameters}. This robustness check is important as these numbers
appear in the exponent which sensitively determines the capture capability. If we divided the evolution of monopole yield only
into two stages, then although we may still get a small reduction exponent at the reference point, it would be hard to argue
that it is insensitive to parameter variations.

We may divide the evolution of the monopole yield into even more stages and obtain more stringent upper bounds on the total reduction exponent for non-diffusive gravitational capture. If the number of stages is large, the discrete sum can be turned into a continuous integral. Let us perform the analysis of the continuous generalization for
\begin{align}
z=z_{\text{mid}1}\equiv 10^{-4}\rightarrow z=z_\text{min}\equiv\frac{T_\text{BBN}}{T_c},
\label{eqn:zstage}
\end{align}
which corresponds to Stage II and III previously. Now we divide Eq.~\eqref{eqn:zstage} into a large number of stages
so that we may write the total reduction exponent (denoted $\mathcal{R}$) as a continuous integral
\begin{align}
\mathcal{R}=2\int_{z_{\text{mid}1}}^{z_\text{min}} J_Fdz^{-1/2}=\sqrt{3}\pi\delta^{1/2}C^{-2}K^{-1}K_1\int_{z_\text{min}}^{z_{\text{mid}1}} y_m^{-1}z^{-3/2}dz.
\end{align}
In analogy to the expressions of $y_{m2}$ and $y_{m3}$ in Eq.~\eqref{eqn:ym123}, here we should write
\begin{align}
y_m=\frac{1}{2}\varepsilon^{-2/3}\gamma^{1/3}K^{-1}C^{4/3}\delta^{2/3}\Big(\frac{T_\text{RH}^\text{max}}{\MPl}\Big)^{-2/3}z^{-2/3}.
\end{align}
Then it is easy to deduce
\begin{align}
\mathcal{R}\simeq 2\sqrt{3}\pi\delta^{-1/6}C^{-10/3}K_1\varepsilon^{2/3}\gamma^{-1/3}
\Big(\frac{T_\text{RH}^\text{max}}{\MPl}\Big)^{2/3}z_{\text{mid}1},
\end{align}
which reads at the reference point
\begin{align}
\mathcal{R}\simeq 1.3\times 10^{-10}.
\end{align}
This is about five orders of magnitude smaller than the estimate of the reduction exponent based on the three-stage expanding/shrinking analysis. The result confirms that the effect of non-diffusive gravitational capture by PBHs
on the monopole yield is tiny.

In the above analyses of gravitational capture of monopoles by PBHs with an extended PBH mass function,
it seems that we have assumed the capture is all diffusive, or all non-diffusive. The actual case is at any
given temperature, some PBH capture is diffusive while some other PBH capture can be non-diffusive, depending
on the PBH mass. Nevertheless, the validity of our main conclusion that both diffusive and non-diffusive gravitational capture by PBHs cannot significantly reduce the monopole yield, is not affected. This is simply because we may disregard the
comparison between $r_c^\gc$ and $\ell$ which is used to distinguish the diffusive and non-diffusive regimes and include
both contributions mathematically. This can only overestimate the reduction of the monopole yield. Such an overestimate
can be divided into a diffusive part and a non-diffusive part which according to our previous analyses neither can reduce
significantly the monopole yield.

We also comment that the neglect of monopole annihilation term in the above analyses also does not affect the main conclusions.
If the initial yield $r_i$ is larger than $r_\star$, monopole annihilation can reduce it to $r_\star$ but not smaller. If the initial yield $r_i$ is smaller than $r_\star$, monopole annihilation cannot reduce it further significantly. This is determined by
the competition between monopole annihilation and cosmic expansion. In Eq.~\eqref{eqn:dlnrdw} the monopole annihilation and
PBH capture contributes independently and whether one term can significantly affect the monopole yield depends on its own competition with the cosmic expansion. Therefore when $r\leq r_\star$ it is not possible to reduce $r$ significantly further
via PBH capture.

\section{Discussion and conclusions}
\label{sec:dnc}

We have generalized the analysis of gravitational capture of magnetic monopoles by PBHs of ref.~\cite{Zhang:2023tfv}
to extended PBH mass functions and different capture types (diffusive and non-diffusive) within the assumption
of radiation domination in the early universe. A general parametrization of the monopole annihilation and capture
term is introduced for solving the evolution of the monopole yield, which is suitable for a variety of scenarios.
We employ the feature of the associated capture term and tricks of inequalities to prove that assuming radiation
domination, gravitational capture of magnetic monopoles of sub-Planckian masses by PBHs cannot significantly reduce the monopole yield beyond the value set by monopole annihilation (or its initial yield if it is smaller). This suggests that
the monopole problem associated with a grand or partially unified gauge theory cannot be solved by PBH capture
in a radiation-dominated era. Or in other words, if we wish to solve the monopole problem by PBH capture, we must
consider matter domination by PBHs. In such a case, consequences of a number of effects must be evaluated, such as
PBH clustering~\cite{Hooper:2020evu} and entropy generation due to PBH evaporation~\cite{Izawa:1984ww}. Moreover, residual magnetic charge fluctuation and ``hot spot'' effects~\cite{Das:2021wei,He:2022wwy} must also be evaluated. Even if PBHs evaporate
before BBN, their abundance is constrained through the associated effect of induced gravitational waves~\cite{Papanikolaou:2020qtd,Papanikolaou:2022chm}, limiting their ability of gravitational capture even in the matter domination period.

An interesting issue related to the gravitational capture of magnetic monopoles by PBHs in the early universe is the formation
of magnetic black holes. Near-extremal magnetic black holes have fascinating theoretical and observational properties, which
have been a subject of intense studies~\cite{Maldacena:2020skw,Bai:2020spd,Liu:2020vsy,Ghosh:2020tdu,Liu:2020bag,Bai:2020ezy,Diamond:2021scl,Chen:2022qvg}\footnote{Properties of PBHs having gravitomagnetic monopole charge are also investigated in the literature; see e.g.~\cite{Chakraborty:2022ltc}.}. In
ref.~\cite{Zhang:2023tfv} we have demonstrated that cosmologically long-lived near-extremal magnetic black holes cannot form from magnetic
charge fluctuation during the gradual diffusive PBH capture process. Due to the inefficiency of non-diffusive gravitational
capture as demonstrated in Sec.~\ref{sec:dnd}, we do not expect non-diffusive PBH capture could lead to cosmologically long-lived near-extremal magnetic black holes. Instead, as shown in ref.~\cite{Zhang:2023tfv}, they can form from magnetic charge fluctuation at PBH formation, when magnetic monopoles inside a horizon volume are collapsed into a black hole almost instantaneously\footnote{This is similar to the formation mechanism of dark extremal black holes studied in ref.~\cite{Bai:2019zcd}.}.
Nevertheless, this formation mechanism entails a monopole yield that is many orders of magnitude larger than
the value allowed by the Parker bound~\cite{Parker:1970xv,Rephaeli:1982nv,Adams:1993fj,Lewis:1999zm,
Medvedev:2017jdn,Kobayashi:2021des,Kobayashi:2022qpl,Kobayashi:2023ryr}. Therefore, in order to have an abundance of near-extremal
magnetic black holes that is of observational interest, some non-inflationary solution to the monopole problem is needed
to get rid of the excessive monopoles. The present study implies that PBH capture in a radiation-dominated era cannot be such
a solution.

The analyses presented in this work can be generalized to studying PBH capture of hidden sector monopoles which
are in thermal equilibrium with the hidden sector plasma, or other SMPs in the diffusive or
non-diffusive regime, which we leave for future work. These studies will help to understand the effect of PBHs on
relic abundance of interesting cosmological relics and clarify the role played by PBHs in the early universe.

\begin{acknowledgments}
Chen Zhang would like to thank Yi-Lei Tang and Sai Wang for helpful discussion. This work was supported by
the National Natural Science Foundation of China (Grants Nos. 11975072 and 11835009)
and the National SKA Program of China (Grants Nos. 2022SKA0110200 and 2022SKA0110203).
\end{acknowledgments}

\bibliography{mmpbhea_v4}
\bibliographystyle{JHEP}

\end{document}